\documentclass[prb,preprint,preprintnumbers,amsmath,amssymb]{revtex4}

\usepackage{graphicx}
\usepackage{amsmath}
\usepackage{amsbsy}
\usepackage{dcolumn}
\usepackage{bm}

\begin{document}

\title{\flushleft{\Huge\sf Electric-field controlled spin reversal in a quantum dot with
ferromagnetic contacts}}

\author{\begin{flushleft}{\sf J.R. HAUPTMANN$^{1\ast}$, J. PAASKE$^{1}$ AND
P.E. LINDELOF$^{1}$}\end{flushleft}} \affiliation{\begin{flushleft}
\footnotesize\normalfont\mbox{\sf $^{1}$The Niels Bohr Institute \&
The Nano-Science Center, University of Copenhagen,
DK-2100 Copenhagen, Denmark}\\
\footnotesize\normalfont\mbox{\sf $^\ast$e-mail:
rahlf@fys.ku.dk.}\end{flushleft}}
\date{\sf\today}

\maketitle


{\bf Manipulation of the spin-states of a quantum dot by purely
electrical means is a highly desirable property of fundamental
importance for the development of spintronic devices such as
spin-filters, spin-transistors and single-spin memory as well as for
solid-state qubits~\cite{ref:Wolf01,ref:Awschalom07,ref:Ohno00,
ref:Recher00,ref:Folk03,ref:Elzerman04}. An electrically gated
quantum dot in the Coulomb blockade regime can be tuned to hold a
single unpaired spin-$1/2$, which is routinely spin-polarized by an
applied magnetic field\cite{ref:Lindelof02}. Using ferromagnetic
electrodes, however, the properties of the quantum dot become
directly spin-dependent and it has been demonstrated that the
ferromagnetic electrodes induce a local exchange-field which
polarizes the localized spin in the absence of any external
fields~\cite{ref:Ralph04,ref:Martinek03a}. Here we report on the
experimental realization of this tunneling-induced spin-splitting in
a carbon nanotube quantum dot coupled to ferromagnetic
nickel-electrodes. We study the intermediate coupling regime in
which single-electron states remain well defined, but with
sufficiently good tunnel-contacts to give rise to a sizable
exchange-field. Since charge transport in this regime is dominated
by the Kondo-effect, we can utilize this sharp many-body resonance
to read off the local spin-polarization from the measured
bias-spectroscopy. We show that the exchange-field can be
compensated by an external magnetic field, thus restoring a
zero-bias Kondo-resonance~\cite{ref:Martinek03a}, and we demonstrate
that the exchange-field itself, and hence the local
spin-polarization, can be tuned and reversed merely by tuning the
gate-voltage~\cite{ref:Martinek05,ref:Sindel07}. This demonstrates a
very direct electrical control over the spin-state of a quantum dot
which, in contrast to an applied magnetic field, allows for rapid
spin-reversal with a very localized addressing.}


Since the discovery of carbon nanotubes they have been intensively
studied for their unique electrical properties. Their high
Fermi-velocity and small spin-orbit coupling make them particularly
well-suited for spintronics applications utilizing the
transformation of spin-information into electrical signals.
Spin-valve effects in nano-junctions with a carbon nanotube (CNT)
spanning two ferromagnetic electrodes have already been
observed~\cite{ref:Sahoo05,ref:Jensen05,ref:Hueso07}, and a strong
gate-dependence of the tunnel magneto-resistance has been
demonstrated for carbon nanotube quantum dots in both the Coulomb
blockade, and the Fabry-Perot regime~\cite{ref:Sahoo05}.

In the intermediate coupling regime, odd numbered quantum dots
exhibit the Kondo effect seen as a pronounced zero-bias conductance
peak at temperatures below a characteristic Kondo temperature,
$T_K$~\cite{ref:Glazman88,ref:Goldhaber98,ref:Lindelof00}. This
effect relies on the conduction electrons being able to flip the
spin of the dot during successive cotunneling-events and is
therefore expected to be sensitive to spin-polarization of the
electrodes. As pointed out by Martinek {et
al.}~\cite{ref:Martinek03a}, quantum charge-fluctuations of the dot
will renormalize the single-particle energy-levels in a
spin-dependent manner and thereby break the spin-degeneracy on the
dot, causing the zero-bias Kondo peak to split in two. This
tunneling-induced exchange-field splitting has since been seen by
Pasupathy {\it et al.}~\cite{ref:Ralph04} in an electromigrated
Ni-gap holding a $C_{60}$-molecule.

A singly occupied level residing just below the Fermi-energy of the
leads is strongly shifted by virtual tunneling {\it out of} the dot,
whereas a level deep below the Fermi-energy (by almost the
charging-energy) is shifted by tunneling of electrons {\it into} the
dot. For spin-polarized electrodes having a difference in the
density of spin-up, and spin-down states, this implies a
spin-splitting of the dot level where sign and size depends on the
applied gate-voltage, i.e. the position of the level below the
Fermi-energy. At the particle-hole symmetric point right between the
empty and doubly occupied states the spin-degeneracy is expected to
be intact~\cite{ref:Choi04}. In a material like Ni, however, one
expects the band structure to be energy dependent and to have a
Stoner splitting, this will break the particle-hole symmetry and
therefore shift the spin-degeneracy point away from the middle of
the diamond~\cite{ref:Martinek05,ref:Sindel07}. Basically all the
theoretical predictions for this pronounced gate-dependence of the
local spin-states are experimentally verified by the
transport-measurements presented below.

\begin{figure}[t]
\includegraphics[width=0.5\columnwidth]{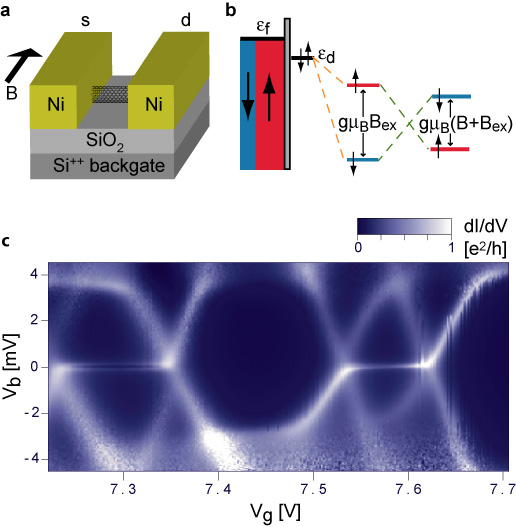}
\caption{\label{fig:figure1}Carbon nanotube quantum dot with
magnetic contacts. \textsf{\textbf{a,}} Schematic view of the device
comprised of a segment of carbon-nanotube on a Si/SiO$_2$ substrate,
contacted by two Ni-leads in a standard field-effect configuration.
A magnetic field is applied along the leads in the plane of the
substrate. \textsf{\textbf{b,}} The local spin-states are split
partly by the tunneling-induced exchange-field, $B_{ex}$ (in orange)
that depends on gate voltage and partly by the external magnetic
field $B$ (in green) resulting from the applied field $B$.
\textsf{\textbf{c,}} Plot of the differential conductance $dI/dV$ as
a function of gate $V_g$, and bias-voltage $V_b$ measured on device
2. A series of 4 Coulomb diamonds can be observed with a pronounced
zero-bias spin-$1/2$ Kondo peak in the two diamonds corresponding to
odd-numbered occupation.}
\end{figure}

The CNTs were grown by chemical vapor deposition on a SiO$_2$ wafer
with a highly doped Si back-gate. The ferromagnetic leads were made
of pure Ni in strips of thickness $\sim$60\,nm, widths
$\sim$300-1000\,nm and separated by
 200-400\,nm. All measurements were made in a
$^3$He/$^4$He dilution refrigerator, with a base-temperature of
30\,mK, corresponding to a minimum electron temperature of 80\,mK,
and using a standard AC-setup with asymmetric bias. A magnetic field
was applied in the direction of the Ni-leads in the plane of the
electrodes as shown in figure~\ref{fig:figure1}\textsf{\textbf{a}}.
We observe a clear even-odd effect with a zero-bias anomaly in every
second Coulomb diamond indicating a spin-1/2 Kondo effect with a
typical Kondo temperature, $T_{K}\sim 1$\,K. Many of the observed
Kondo anomalies showed a gate-dependent splitting and here we
discuss two different devices for which this dependence is
particularly clear, in the supplement two other examples are shown.

\begin{figure}[t]
\includegraphics[width=0.5\columnwidth]{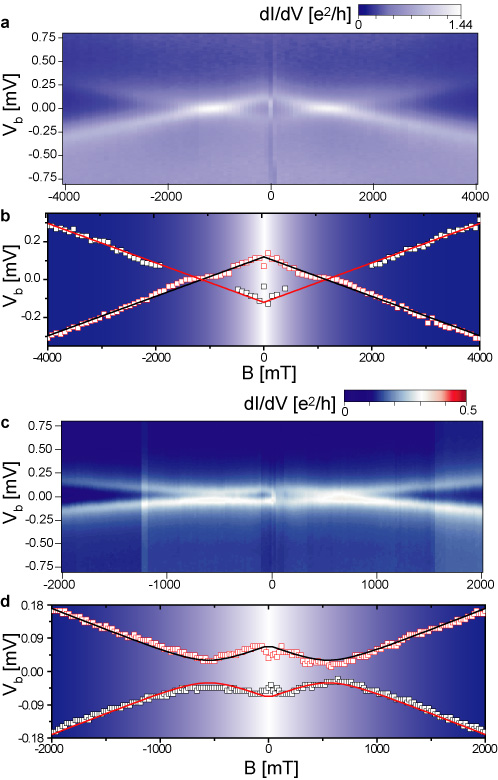}
\caption{\label{fig:figure2}Kondo peak splitting as a function of
applied magnetic field at fixed gate-voltage. \textsf{\textbf{a,}}
$dI/dV$ vs. bias-voltage and magnetic field measured on device\,1,
sweeping from -4\,T to 4\,T. The plot is recorded in the middle of
an odd numbered spin-1/2 Kondo diamond. The observed splitting of
the Kondo peak is due to the the exchange field $B_{ex}$ and
external field $B$. At $B\sim\pm1.12\,$T the external field is
compensating the exchange-field and restoring the zero-bias Kondo
peak. A sudden change in $dI/dV$ is observed at $B\sim-80\,$mT due
to a switching of one of the two contact-domains. At $B\sim80\,$mT
the other domain has flipped and we are back in a parallel
configuration. \textsf{\textbf{b,}} Plot of the peak positions from
\textsf{\textbf{a,}} offset with $V_b=5\,\mu$V to symmetrize the
plot. The black, and red lines are fits to the peak positions and
have a slope of $\pm1.04*10^{-4}\,$V/T corresponding to $g\sim1.8$.
\textsf{\textbf{c,}} Plot as in \textsf{\textbf{a,}} measured on
device\,2, sweeping from -2\,T to 2\,T. In this device the
exchange-field is never completely compensated, indicating a
misalignment of the domain magnetization direction and the applied
magnetic field, ${\bf B}$. A domain-switch causes a sudden jump in
the splitting at $B\lesssim |40|\,$mT. \textsf{\textbf{d,}} Peak
positions from \textsf{\textbf{c,}} offset by $V_b=12\,\mu$V. The
black, and red lines are fits to the plots where the domain
magnetization direction is at an angle of 25$^\circ$ to the external
field (see text).}
\end{figure}

Magnetic force microscopy images of devices similar to the ones
measured indicate that the CNT quantum dot is most likely coupled to
one single domain in both source, and drain electrode. Applying a
strong magnetic field serves partly to align the two contact-domains
and partly to provide a Zeeman-splitting of the local spin.
Figure~\ref{fig:figure2} shows the conductance vs. bias-voltage and
external field, measured for a gate-voltage tuned to the middle of
an odd occupied Coulomb blockade valley in the Kondo regime.
Device\,1 (figure~\ref{fig:figure2}\textsf{\textbf{a,b}}) shows a
simple linear behavior in which the single-domain magnetization and
hence the exchange field, ${\bf B}_{ex}$, is aligned with the
external field, ${\bf B}$. The exchange field can therefore be
completely compensated and the zero-bias Kondo peak is seen to be
restored at $B\sim\pm1.12\,$T, giving an indirect measure of the
exchange-field at this gate-voltage. In device\,2
(figure~\ref{fig:figure2}\textsf{\textbf{c,d}}), on the other hand,
the Kondo peak is never fully restored and the splitting merely
reaches a minimum at $B\sim\pm0.6\,$T. This indicates that the
exchange-field lies at an angle to the external field, and fitting
the B-dependence of the peak-splitting by $e\Delta V=g\mu_{B}|{\bf
B}+{\bf B}_{ex}|$, we find this angle to be $\angle({\bf B},{\bf
B}_{ex})\sim 25^\circ$(cf. supplementary material).

Figure~\ref{fig:figure2} was recorded by sweeping the field from
large negative to large positive values, a sudden decrease(increase)
of the splitting is clearly visible at small negative(positive)
fields. A similar switching was observed in
Ref.~\onlinecite{ref:Ralph04} and can be ascribed to a sudden
switching from parallel (P) to anti-parallel (AP) configuration of
the contact-domains driven by domain interactions (cf. supplementary
material). In the AP-configuration the tunnel-induced exchange-field
is nearly canceled, unless the couplings to source, and
drain-electrodes are very different. In terms of the
conduction-electron spin-polarization,
$P=(\nu_{F\uparrow}-\nu_{F\downarrow})/\nu_{F}$, and the
tunnel-broadening, $\Gamma_{s(d)}=\pi\nu_{F}|t_{s(d)}|^{2}$, where
$\nu_{F}=\nu_{F\uparrow}+\nu_{F\downarrow}$ denotes the density of
states (DOS) in the leads and $t_{s(d)}$ is the tunneling amplitude
to source(drain), one expects~\cite{ref:Ralph04,ref:Utsumi05} that
$g\mu_BB^{(AP/P)}_{ex}=\Delta\varepsilon^{(AP/P)}_{ex}=aP(\Gamma_{s}\mp\Gamma_{d})$.
Comparing now the exchange-fields in P, and AP-configuration we
deduce that $\Gamma_{s}/\Gamma_{d}\sim 3$, this corresponds to a
zero-bias Kondo peak height of
$(2e^{2}/h)4\Gamma_{s}\Gamma_{d}/(\Gamma_{s}+\Gamma_{d})^{2}=1.5\,
e^{2}/h$ roughly consistent to the value of $1.4 \,e^{2}/h$,
measured on device\,1 in the middle of the Coulomb blockade diamond
for a fully compensating magnetic field near $B=\pm$1.12 T.

\begin{figure}[b]
\includegraphics[width=\columnwidth]{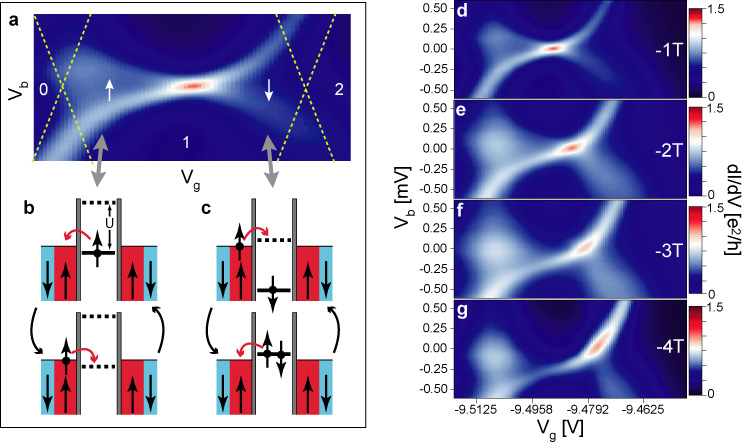}
\caption{\label{fig:figure3}Gate dependence of the spin-states
probed by the Kondo effect. \textsf{\textbf{a,}} Plot from
\textsf{\textbf{d,}} with yellow dotted lines added to indicate the
diamond edges. Numbers indicate the number of electrons in one
orbital state. Arrows indicates the local spin ground-state which
changes with the exchange-field as the gate-voltage is varied. These
spin ground-states on the dot are consistent with the observation in
\textsf{\textbf{d-g,}} that the conductance peak (red spot) moves to
the right as the field is increased. \textsf{\textbf{b, c,}}
Illustration of the virtual tunneling-processes leading to a
spin-up(down) ground-state in the left(right) side of the dot, the
full drawn lines are filled states and the dotted lines are empty
states. Upper and lower diagrams correspond to ground, and excited
states. Red spin-up represents the majority s-electron-band in the
leads, hybridizing the most with the dot-electrons.
\textsf{\textbf{d-g,}} Plots of $dI/dV$ as a function of gate, and
bias-voltage, measured on device\,1 in different external magnetic
fields. The Kondo peak is clearly seen to have a gate-dependent
splitting. Full compensation in the middle of the diamond is
obtained close to $B=-1$T as seen in \textsf{\textbf{d}}.}
\end{figure}

In a Coulomb blockade diamond with an odd number of electrons, the
left, and right charge-degeneracy points correspond to respectively
emptying the dot or filling it by one extra electron. With a finite
spin-polarization in the leads, tunneling of majority-spins, spin-up
say, will be favored by the higher-density of states and a dot-state
of spin-up will therefore be shifted further down in energy than a
spin-down state, as long as one is closer to the left hand side
(l.h.s.) of the diamond. Due to the Pauli-principle, the
majority-spins can only tunnel into the dot if the residing electron
is in a spin-down state. It is therefore the spin down state which
is lowered the most near the right hand side (r.h.s.) of the
diamond. This simple mechanism of level-renormalization, illustrated
in figure~\ref{fig:figure3}\textsf{\textbf{a}}-\textsf{\textbf{c}},
is encoded in the exchange-field, given to good accuracy by
\begin{equation}
\Delta\varepsilon_{ex}(\varepsilon_d)=e\Delta_{st}
+(P\Gamma/\pi)\ln(|\varepsilon_{d}|/|U+\varepsilon_{d}|)
\end{equation}
to leading order in the tunneling-amplitude~\cite{ref:Sindel07} and
assuming a constant density of states. $U$ is the charging-energy
and $\varepsilon_{d}$ is the dot-level-position which is
proportional to the gate-voltage. Notice the strong negative and
positive logarithmic corrections for $\varepsilon_{d}$ close to $0$
or $-U$, respectively (corresponding to left, and right borders of
the diamond). In the middle of the diamond, $\varepsilon_{d}=-U/2$,
the exchange-field is zero except for a constant term,
$\Delta_{st}$, which reflects a Stoner-splitting between the spin-up
and spin-down bands~\cite{ref:Martinek05,ref:Sindel07}. Notice that
possible stray-fields from the magnetic contacts would also
contribute with a gate-independent spin-splitting (cf. supplementary
material).

Depending on the magnitude of $\Delta_{st}$, Equation~(1) predicts a
change of the ground state spin direction as the localized level is
moved from $\varepsilon_d=0$ to $\varepsilon_d=-U/2$. Such a spin
reversal can be observed in
figure~\ref{fig:figure3}\textbf{\textsf{d}}-\textbf{\textsf{g}} and
occurs where the Kondo ridges cross and the spin states are
degenerate\cite{ref:footnote}. The movement of the conductance peak
(red dot) as function of magnetic field, seen in
figure~\ref{fig:figure3}\textsf{\textbf{d}}-\textsf{\textbf{g}},
confirms that the ground-state spin can indeed be reversed by
changing the gate-voltage. Furthermore, the direction of the motion
shows that it is predominantly spin-up (i.e. the spin along the
external magnetic field) electrons which tunnel on and off the dot,
giving a ground-state with spin-up at the l.h.s of the diamond and
spin-down at the r.h.s. This confirms earlier observations in
conventional superconducting
tunnel-junctions~\cite{ref:Tedrow71,ref:Kim04} indicating that the
tunneling electrons in Ni have majority-spin, i.e. positive
polarization. Most likely, the more mobile $s$-electrons taking part
in the tunneling are spin-polarized by their hybridization with the
localized $d$-electrons~\cite{ref:Tsymbal97,ref:Mazin99}.

\begin{figure}[t]
\includegraphics[width=0.5\columnwidth]{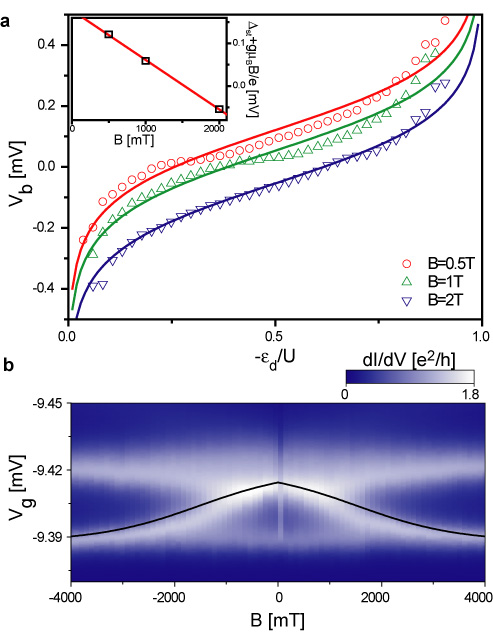}
\caption{\label{fig:figure4}Gate dependence of the exchange-field
for different applied magnetic fields. \textsf{\textbf{a,}}
Scatter-plot of the dominant Kondo peak position, in bias-voltage,
as a function of level-position $\varepsilon_d(\propto V_g)$
gate-voltage for device\,1. Peak-positions are read off from plots
similar to those in figure 3\textsf{\textbf{d-g}} (cf. supplementary
material). The different colors correspond to different magnetic
fields. The gate-axis has been displaced and normalized with the
charging energy such that the dot is emptied at $\varepsilon_{d}=0$
and becomes doubly occupied at $\varepsilon_{d}=-U$. Full lines are
fits to the data, using $(\Delta\varepsilon_{ex}-g\mu_{B}B)/e$
defined by Eq.(1) and with $g=2$. All three curves are fitted by
$P\Gamma=0.113*\pi$. The inset shows a fit to the gate-independent
term $\Delta_{st}-g\mu_{B}B/e$ with $\Delta_{st}=0.18$\,mV.
\textsf{\textbf{b,}} $dI/dV$ as a function of gate-voltage and
magnetic field measured at zero bias voltage. The two nearly
horizontal lines are Coulomb peaks and the white line moving from
the lower to the upper peak as $B$ is lowered, is a Kondo peak, thus
mapping out where the spin states are degenerate. The black line is
described by the function
$\varepsilon_{d}=U/(e^{(e\Delta_{st}-g\mu_{B}B)\pi/P\Gamma}+1)-c$
where $c$ is a constant determined by the first Coulomb peak and $U$
is the charging energy. $c$ and $U$ are read off from the plot while
the other constants are determined from the fits in
\textsf{\textbf{a}}. Notice that, like in figure \ref{fig:figure2},
a domain switch is seen at $|B|\lesssim80\,$mT.}
\end{figure}

In order to substantiate the detailed gate-voltage dependence of the
exchange-field, we have extracted the peak-positions from plots like
those in
figures~\ref{fig:figure3}\textsf{\textbf{d}}-\textsf{\textbf{g}} and
fitted them by $(\Delta\varepsilon_{ex}(\varepsilon_d)-g\mu_BB)/e$
with $\Delta\varepsilon_{ex}$ given by equation~(1). The result is
shown in figure~\ref{fig:figure4}\textsf{\textbf{a}} where the inset
shows $\Delta_{st}-g\mu_BB/e$ as a function of $B$. The slope of the
line corresponds to $g=2$ as expected for a carbon nanotube. The
white ridge in Figure~\ref{fig:figure4}\textsf{\textbf{b}} traces
the magnetic field dependence of the gate-voltage at which the spin
states are degenerate, showing a gratifying resemblance to the
theory-plots presented in
Refs.~\onlinecite{ref:Martinek05,ref:Sindel07}. Inverting
equation~(1) and using the fitting-parameters deduced from
figure~\ref{fig:figure4}\textsf{\textbf{a}}, we arrive at the
parameter free fit seen as the black line in
figure~\ref{fig:figure4}\textsf{\textbf{b}}.




\vspace*{8mm} \noindent {\bf\sf Acknowledgements}

\noindent {\scriptsize We thank R.~Gunnarsson, L.~Borda,
K.~Flensberg, P.~Hedeg{\aa}rd, and C.~M.~Marcus for discussions. For
financial support we thank Danish Agency for Science Technology and
Innovation (FTP 274-05-0178) and EU projects SECOQC and CARDEQ.}

\vspace*{4mm} \noindent {\bf\sf Competing financial interests}

\noindent {\scriptsize The authors declare that they have no
competing financial interests.}


\newpage

\cleardoublepage

\begin{flushleft}
\section*{\Large{Supplementary material for "Electric-field controlled spin
reversal in a quantum dot with ferromagnetic contacts"}}
\end{flushleft}

\section{Domain switching and stray-fields}

Magnetic force microscopy images of devices similar to the ones
measured, reveal a typical domain-size of
$V_{dom}=100\times500\times50\,nm^3$, which suggests that the carbon
nanotube quantum dot is most likely coupled to one single domain in
both source, and drain electrodes. Using the saturation
magnetization for Ni, $M_{s}=0.61{\rm T}/\mu_{0}$, the magnetic
moment of a single contact-domain can be estimated by
$V_{dom}M_{s}$. With 200\,nm between the electrodes, this gives a
crude estimate of the dipole-dipole interaction energy of the two
contact-domains with an energy-gain of roughly 200\,eV between the
parallel (P) and anti-parallel (AP) configurations. Starting in the
P-configuration for a large applied magnetic field, this in turn
implies a transition to AP-configuration once the external field
becomes smaller than some 30\,mT. The switching from P to AP
observed in article-figures~2\textsf{\textbf{a,c}} occurs at fields
smaller than 80\,mT for device\,1 and close to 40\,mT for device\,2.
Article-figures~2\textsf{\textbf{a,c}} were recorded with a
resolution in $B$ of 80\,mT and 10\,mT, respectively. This simple
estimate of course neglects all intra-lead domain-interactions,
which cannot a priori be assumed to be less important.

From this we can also estimate the stray-field at the middle between
the two electrodes. In the AP-configuration it is close to zero and
in the P-configuration it must be roughly $B_{stray}\sim 100$\,mT,
i.e. a factor of 10 and 5, respectively, smaller than the
tunneling-induced exchange-fields which we read off from
article-figures~2\textsf{\textbf{a,c}}. We note that this rather
crude estimate is comparable in magnitude to the estimates made in
Refs.~\onlinecite{ref:sRalph04,ref:sSahoo05,ref:sMeier06}. Notice
also that stray-fields cannot account for the gate dependence
observed in article-figures~3\textsf{\textbf{d-g}}.

The fact that the exchange-field cannot be completely compensated by
the external field in device\,2 is interpreted as a misalignment of
${\bf B}_{ex}$ and ${\bf B}$. Since bulk Ni has only weak magnetic
anisotropy, this is most likely due to the reduced geometry and
surface roughness of the leads. The fit shown in
article-figure~2\textsf{\textbf{d}} takes a fixed direction of the
magnetization which is found to be $\angle({\bf B},{\bf B}_{ex})\sim
25^\circ$. We have also tried to make another fit with a finite
anisotropy barrier, using a simple Stoner-Wohlfarth model where the
magnetic energy of the system is given by
\begin{equation}
E=K\sin^{2}(\theta-\phi)-BM\cos(\phi),
\end{equation}
in terms of applied field $B$ and magnetization $M$ at a relative
angle $\phi$, together with an angle $\theta$ between an easy-axis
and the applied field. $K$ parameterizes the anisotropy barrier,
thus allowing the magnetization to align with ${\bf B}$ for large
enough values of $BM/K$. Minimizing this energy as a function of
angle for a given applied field gives a hysteretic magnetization
curve from which we can then infer the sum of applied, and
exchange-field to determine the spin-splitting as a function of $B$.
Nevertheless, this fit performed no better than the far simpler fit
to a fixed angle.

\section{Additional plots for devices 1 and 2}
\subsection{Device\,1}
\begin{figure}
    \includegraphics[width=0.95\columnwidth]{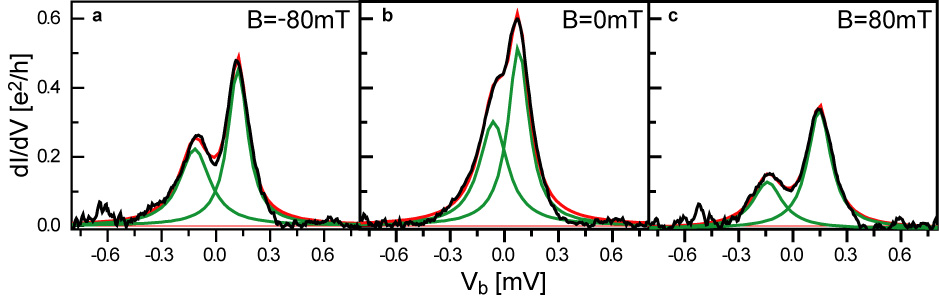}
    \caption{\label{fig:supp1} Plots of dI/dV as a function of bias
    voltage
    for $B=-80\,$mT, $B=0\,$mT and $B=80\,$mT the plots are
    line-cuts through the plot shown in
    article-figure~1\textsf{\textbf{a}},
    measured on device\,1. The black curves are
    the measurements with a baseline subtracted. The green lines are
    Lorentzian fits to the peaks and the red lines are the resulting
    double-peak fits. The center-to-center distance of the Lorentzians in each
    of the three plots gives the splitting of the spin-up and spin-down
    states on the dot. The splittings are $0.239$, $0.139$ and
    $0.287\,$mV for \textsf{\textbf{a}}, \textsf{\textbf{b}} and
    \textsf{\textbf{c}}, respectively, corresponding to domain
    configurations which are \textsf{\textbf{a}} parallel,
    \textsf{\textbf{b}} anti-parallel and {\textsf{\textbf c}}
    parallel.}
\end{figure}
The ratio of the couplings to respectively source and drain
electrodes can be estimated from the formula
\begin{equation}
    \frac{\Gamma_s}{\Gamma_d}= \frac{\Delta V_P-2g\mu_BB/e+\Delta
    V_{AP}}{\Delta V_P-2g\mu_BB/e-\Delta V_{AP}},
\end{equation}
where $V_P$($V_{AP}$) denotes the splittings in the parallel
(antiparallel) configuration which are read off from
figure~\ref{fig:supp1}. This gives the estimate
$\Gamma_s/\Gamma_d\sim 3$, cited in the main paper.

\begin{figure}
    \includegraphics[width=0.45\columnwidth]{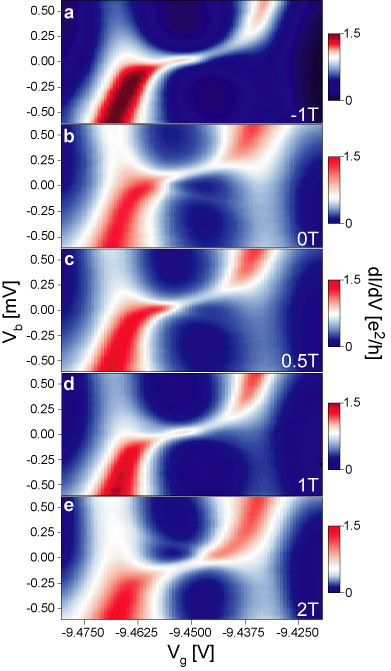}
    \caption{\label{fig:K12postswitch} Gate-dependence of the
    spin-states probed by the split Kondo-peak measured on device\,1
    post switch. \textsf{\textbf{a-e}} show the differential
    conductance as a function of gate, and bias-voltage measured at
    different magnetic fields. Notice that the crossings of the
    Kondo ridges move to the right as the magnetic field is
    increased. It is from these plots the scatters and the fits in
    article-figure~4~\textsf{\textbf{a}} is made.}
\end{figure}

The plots in figure~\ref{fig:K12postswitch} shows the differential
conduction as a function of gate and bias voltage measured at
different magnetic fields. The measurements are made on device\,1
after a switch, i.e. a sudden, permanent change in the device
behavior. Notice that the crossing of the Kondo ridges moves from
left to right as the magnetic field is increased. It is these
measurements that the scatter and fits in
article-figure~4\textsf{\textbf{a}} are made from. Note that the
plot measured at $B=-1\,$T has also been fitted but is not shown in
article-figure~4\textsf{\textbf{a}} since it is placed right on top
of $B=1\,$T. The plot made at $B=0\,$T is not included in the
fitting since the domain configuration is expected to be different.
The point corresponding to $B=1$\,T in the inset of
article-figure~4\textsf{\textbf{a}} is really the average of the
fitted constants, 0.065~mV and 0.053~mV, obtained at $B=-1$\,T and
$B=1$\,T, respectively.

\subsection{Device\,2}
\begin{figure}
    \includegraphics[width=0.45\columnwidth]{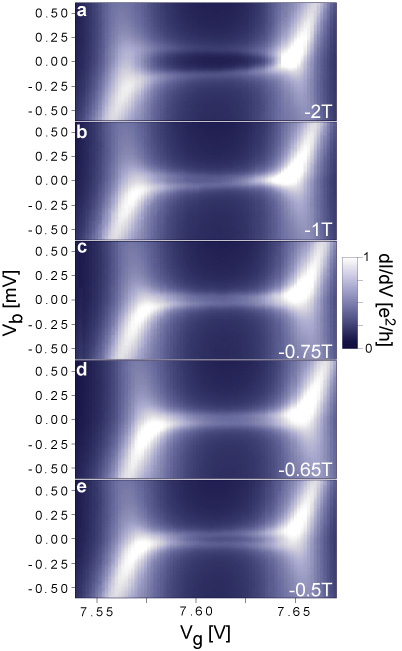}
    \caption{\label{fig:K3} Gate dependence of the spin-states
    probed by the Kondo effect measured on device\,2.
    \textsf{\textbf{a-e}} show the differential conduction as a
    function of gate and bias voltage measured at different magnetic
    fields. Notice that the Kondo ridges do not cross in any of
    the plots. This is interpreted as a misalignment of the domains and the
    external field.}
\end{figure}
The plot shown in figure~\ref{fig:K3} shows the differential
conductance as a function of gate, and bias voltage measured at
different magnetic fields. The slopes of the Kondo ridges in
\textsf{\textbf{a, b}} and in \textsf{\textbf{e}} are clearly
opposite, although the measurements at intermediate magnetic fields
(\textsf{\textbf{c, d}}) show no crossing of the Kondo ridges. We
ascribe this to a misalignment of the domain magnetization direction
(and therefore the exchange field) and the external magnetic field.

\section{Plots for devices 3 and 4}
\begin{figure}
    \includegraphics[width=0.45\columnwidth]{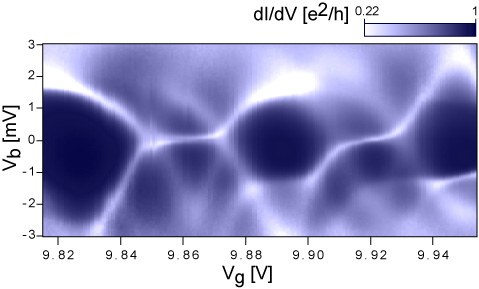}
    \caption{\label{fig:K5} Five Coulomb diamonds with a Kondo resonance
    in every second. The plot show the differential conduction as a function
    of gate and bias voltage.}
\end{figure}
Figure~\ref{fig:K5} shows the differential conduction as a function
of gate, and bias voltage. Two Kondo ridges near zero bias can be
observed and henceforth we shall refer to the corresponding
odd-numbered Coulomb-blockade diamonds as devices 3 and 4,
respectively. These measurements are made on the same sample as
device 1 but off-set in gate voltage by roughly 20~V.

\begin{figure}
    \includegraphics[width=0.45\columnwidth]{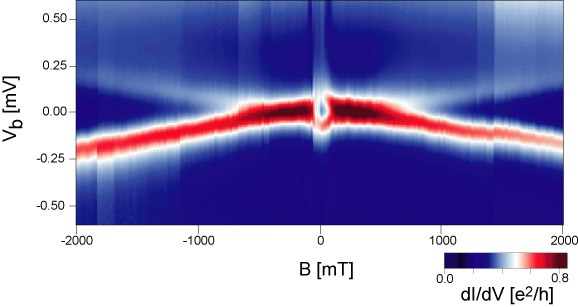}
    \caption{\label{fig:K5-b-Vb} Kondo peak splitting as a function
    of magnetic field at fixed gate voltage measured on device\,3. The
    plots shows the differential conduction as a function of bias and magnetic
    field, swept from 2\,T to -2\,T.}
\end{figure}
Figure~\ref{fig:K5-b-Vb} shows $dI/dV$ as a function of bias voltage
and magnetic field. The measurements are made close to the middle of
the Coulomb-blockade diamond in device\,3. A linear splitting of the
Kondo peak with magnetic field can be observed, as would be expected
since this device should be contacted by the same domains as
device\,1 and the domain configuration and magnetization direction
in the leads should be independent of the gate voltage.

\begin{figure}
    \includegraphics[width=\columnwidth]{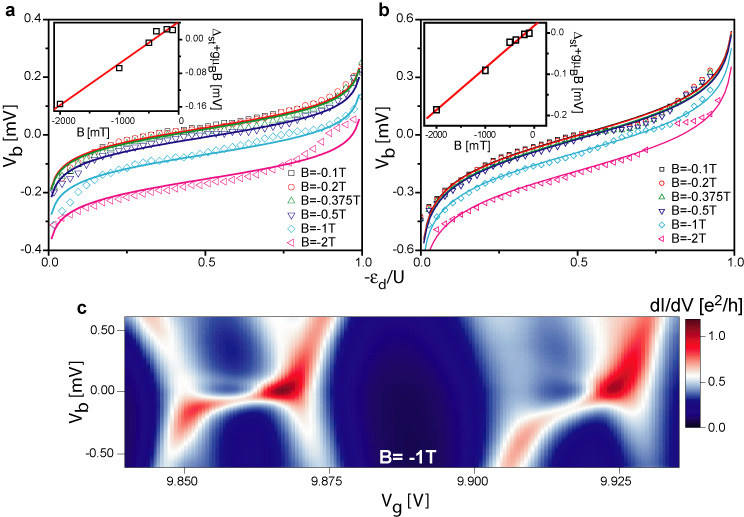}
    \caption{\label{fig:K5-fit} Gate dependence of the exchange
    field for different applied magnetic fields measured on
    \textsf{\textbf{a}} device\,3 and \textsf{\textbf{b}}
    device\,4. {\textsf{\textbf c}} Show the differential
    conduction as a function of bias and gate voltage the
    measurement are made with an external field of -1\,T. Four
    Coulomb peaks and two Kondo resonances are observed the 1st resonance
    corresponds to device\,3 and the 2nd to device\,4. \textsf{\textbf{a, b}}
    shows the scatter plot of the peak position read off from
    measurement like the one shown in \textsf{\textbf{c}}, where
    the gate voltage have been normalized with the charging energy.
    The different colors and shapes corresponds to different
    external magnetic fields. The lines are fits to the scatters.
    The insert shows the fitting constant $\Delta_{st}+g\mu_BB$
    for the different magnetic fields and a linear fit to these
    points.}
\end{figure}
Figure~\ref{fig:K5-fit}\textbf{c} shows a zoom-in on the Kondo
ridges, measured at $B=-1$\,T. A gate dependent splitting of the
Kondo resonance is clearly observed in both device. The different
slopes indicate a change in $\Gamma=\Gamma_s+\Gamma_d$ from one
Kondo resonance too the next, most likely due to a difference in
couplings to the two orbitals in the CNT.

The splittings of the Kondo peaks have been read off and plotted in
figure~\ref{fig:K5-fit}\textsf{\textbf{a,b}}. The lines are fits to
the scatter-plots and the different colors correspond to different
external magnetic fields. The fitted values used for $P\Gamma/\pi$
are 0.045 and 0.117\,meV for the groups of fits in
\textsf{\textbf{a}} and \textsf{\textbf{b}}, respectively. The
insets in both figures show $\Delta_{st}+g\mu_BB$ as function of
$B$. The two linear fits yield a slope of 9.9$\times10^{-5}$
(\textbf{a}) and $1.0\times10^{-4}$ (\textbf{b} ), corresponding to
a $g$-factor of 1.7 and 1.8, respectively.
\begin{figure}
\includegraphics[width=\columnwidth]{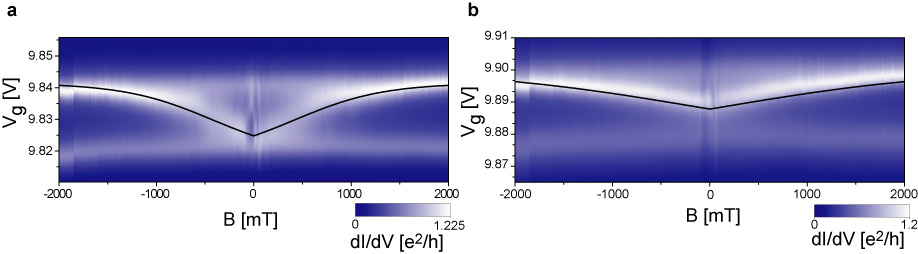}
\caption{\label{fig:K5-vg} Gate dependence of the exchange field at
zero bias. The plots shows the differential conduction as a function
of gate voltage and external magnetic field for
device\,3~\textsf{\textbf{a}} and device\,4~\textsf{\textbf{b}}. The
two horizontal lines in each plot are due to Coulomb peaks. The
white line crossing from one peak to the other \textsf{\textbf{a}}
or halfway between them \textsf{\textbf{b}} are mapping out the
Kondo peak movement (the point where the spin states are degenerate)
as a function of magnetic field. The black line are parameter free
fit where the constant have been found from the fits in
figure~\ref{fig:K5-fit}. Similar to the parameter free fit shown in
article-figure~4\textsf{\textbf{b}}.}
\end{figure}

Figure~\ref{fig:K5-vg} shows $dI/dV$ as function of gate voltage and
magnetic field. Panels \textbf{a} and \textbf{b} correspond to
devices 3 and 4, respectively. The black line is plotted using the
procedure described in the caption of
article-figure~4\textsf{\textbf{b}}, now with parameters obtained
from figure~\ref{fig:K5-fit}.


\end{document}